# The cuprate phase diagram and the influence of nanoscale inhomogeneities.

Nader Zaki, Hongbo Yang, Jon Rameau, Peter D. Johnson

Condensed Matter Physics and Materials Science Department,

Brookhaven National Laboratory,

Upton, NY 11973

Helmut Claus, David G. Hinks

Materials Science Division,

Argonne National Laboratory

Argonne, Illinois, 60439

Abstract:

The phase diagram associated with high $T_c$ superconductors is complicated by an array of different ground states. The parent material represents an antiferromagnetic insulator but with doping superconductivity becomes possible with transition temperatures previously thought unattainable. The underdoped region of the phase diagram is dominated by the so-called pseudogap phenomena whereby in the normal state the system mimics superconductivity in its spectral response but does not show the complete loss of resistivity associated with the superconducting state. An understanding of this regime presents one of the great challenges for the field. In the present study we revisit the structure of the phase diagram as determined in photoemission studies. By careful analysis of the role of nanoscale inhomogeneities in the overdoped region, we are able to more carefully separate out the gaps due to the pseudogap phenomena from the gaps due to the superconducting transition. Within a mean-field description, we are thus able to link the magnitude of the doping dependent pseudogap directly to the Heisenberg exchange interaction term, $J \sum s_i \cdot s_j$, contained in the $t - J$ model. This approach provides a clear indication that the pseudogap is associated with spin singlet formation.

**Introduction:**

The strongly correlated high-temperature cuprate superconductors continue to present challenges for the research community. It is well established that the ground state of the parent materials are antiferromagnetic insulators. Upon doping, long range magnetic order is lost and replaced by superconductivity with transition temperatures previously thought unattainable. However, the phase diagram is also dominated by the so-called pseudogap regime where at high temperatures, the system mimics superconductivity in its spectral properties but only enters the superconducting state at lower temperatures. It is therefore generally thought that a determination of the source of the pseudogap and the subsequent unraveling of the complexities of the cuprate phase diagram will ultimately provide a pathway to a final understanding of the physics of high $T_c$ superconductivity and from that, a possible pathway to even higher superconducting transition temperatures. However questions continue to arise as to whether the pseudogap region reflects preformed pairing of electrons or a competition between different orders inhibiting the development of the superconducting state. An example of the latter is given by charge ordering and superconductivity as found, for instance, in the dichalcogenides.[1] X-ray scattering studies have identified short range charge ordering in the cuprates[2,3] but it appears confined to the region corresponding approximately to the underdoped side of the superconducting dome and certainly at temperatures lower than those associated with the pseudogap at the same doping levels. Long range ordering does appear to compete with the superconductivity as evidenced by the dip in $T_c$ in the vicinity of 1/8 doping, a doping level at which certainly the $La_{0.875}Ba_{0.125}CuO_4$ system is known to exhibit static "stripe order".[4] Before attempting to resolve such issues it is important to first establish the correct form for the phase diagram. Figure 1 shows two frequently presented formats. In (a) the pseudogap line touches the superconducting dome tangentially; in (b) the line penetrates the dome to strike the doping axis at some critical point, which may or may not be quantum critical. Photoemission represents one of the key probes of the low lying excitations and associated gaps in these materials.[5] In presenting the cuprate phase diagram, photoemission-based studies almost universally propose that the temperature-dependent "pseudogap line" does in fact brush the superconducting dome tangentially on the overdoped side, as indicated in figure 1(a).[5,6] However, such a picture is difficult to reconcile with other experiments and models that point to the possibility of a critical point inside the superconducting dome, as in fig. 1(b).

Early angle-resolved photoelectron spectroscopy (ARPES) studies of the pseudogap regime in $Bi_2Sr_2CaCuO_{2+\delta}$ (Bi2212) identified disconnected Fermi arcs in the nodal region reflecting the presence of a spectral gap, the "pseudogap", in the anti-nodal direction, the latter corresponding to the copper-oxygen bond directions.[7,8] Photoemission studies of highly overdoped materials on the other hand found evidence of a full Fermi surface consistent with a more metallic phase.[9] Closed Fermi surfaces are certainly expected for condensed matter systems in general and as such, the Fermi arcs in the underdoped materials have been the subject of considerable investigation. Several studies have been interpreted as indicating a temperature dependent arc length,[10] others a doping dependent length.[11,12] More detailed studies have suggested that the arcs do, in fact, represent one side of an asymmetric hole-pocket[13] consistent with several models of the doped Mott insulator at low doping.[14,15,16] As a function of increased doping, the hole pockets grow with an area proportional to $x$, the doping level, until at some critical doping level the pseudogap disappears[15]; at this level, the pockets switch to the full Fermi surface associated with a more metallic state. The full Fermi surface encloses a hole area equal to $(1 + x)$. These observations have indeed also been made recently using Spectroscopic Imaging Scanning Tunneling Microscopy (SISTM) studies on the Bi2212 system which showed the disconnected Fermi arcs switching to a full Fermi surface at a doping level of approximately $x = 0.19$.[17] Furthermore recent high magnetic field studies of the Hall coefficient in a related cuprate, $YBa_2Cu_3O_y$ (YBCO), also point to a crossover from hole-pockets with area proportional to $x$ in the underdoped phase to the full Fermi surface with area proportional to $(1+x)$ at the same critical doping level.[18] Such a reconstruction of the Fermi surface would imply that there is no pseudogap at doping levels higher than the critical doping level, as suggested in the phase diagram presented in figure 1(b). Other observations including Raman studies pf Bi2212[19] and neutron scattering studies of YBCO[20] indicate a break in behavior in the vicinity of optimal doping or slightly higher. Further, tunneling spectroscopy studies of break junctions on highly overdoped Bi2212 indicate a lack of pseudogap.[21] How then do we reconcile these latter observations with the observation of a gap above Tc in photoemission studies of overdoped Bi2212?

Examination of figure 1(b) indicates an opportunity for studying two particularly interesting regions:(i) lower doping at higher temperatures providing access to the pseudogap interactions without the complications of superconductivity and/or short range charge ordering (lattice disorder), and (ii) higher doping levels above the critical doping level providing access to the superconducting properties alone. Thus in the present study we bring new insights into the discussion by considering the properties of these two distinct regions.

In the overdoped region, consideration of the nanoscale inhomogeneities observed in this system[22] leads to a picture whereby the gap observed above the superconducting transition temperature, $T_c$, appears to be associated entirely with superconductivity in the inhomogeneities. We do not provide any insight into the origin of these inhomogeneities but note that they may reflect either chemical doping or electronic phase separation or indeed both. In the underdoped region, examination of the pseudogap line as a function of doping, $x$, confirms a picture of the phase diagram such that the pseudogap onset $T^*(x)$ line penetrates the superconducting dome and points to a critical doping level, $x_c$, at approximately 0.19. We associate this pseudogap line with the crossover from doped Mott insulator to marginal Fermi liquid or strange metal and the pseudogap itself with the singlets associated with an RVB type of spin liquid.[23]

**Experimental Results:**

a) **The Overdoped Regime:**

We first examine the properties of the electronic structure in the anti-nodal region in Bi2212 at doping levels of 0.2 and above using the technique of angle-resolved photoemission spectroscopy. Fig. 2(a) shows the temperature dependence of the measured photoemission spectra recorded at the antinodal point, indicated in the inset 2(b). To gain a more accurate determination of gap size we show in figure 2(c) the (normalized) spectra symmetrized around the chemical potential, a technique frequently used in a number of earlier studies. Our justification for doing this is shown in the supplementary information where we compare the symmetrized spectrum with that obtained by normalizing the raw data with the appropriate temperature dependent Fermi-Dirac function. There is almost perfect agreement between the two methods for this part of the Fermi surface. We have made a similar observation in the past.[11,13]

We now turn to a slightly different analysis from that used in earlier studies. For each spectrum in fig. 2(c), we fit the structure with two Lorentzian peaks and determine the full width at half maximum (FWHM) across both peaks. With increasing temperature as the gap between the two peaks disappears we eventually obtain the FWHM associated with a single peak. The FWHM determined in this fashion is shown in figure 3(a), where it will be seen that the FWHM initially shows an increase as the gap starts to close and then decreases again before resuming the normal linear temperature dependence that we associate with strange metal or marginal Fermi liquid behavior.[24,25] Figure 3(b) shows the disappearance of the gap between the two fitting Lorenztians. Similar data from other doping levels is presented in the supplementary information. How do we interpret these observations?

Firstly, in the absence of the pseudogap at overdoping, we associate the two peaked structure with the superconducting state and hence Cooper paired electrons. At temperatures close to the gap closing, indicative of the breakup of the Cooper pair, we anticipate an initial broadening of the peak associated with the reduced lifetime of the electron in the paired state. Several earlier studies have indicated that this lifetime is effectively a step function around the transition temperature.[26,27] However, after the gap has closed as indicated in fig 3(b), we note that the width continues to decrease over a finite temperature range, behavior that seems inconsistent simply with the gap closing and a step function associated with lifetime. The solution to this issue comes from SISTM studies of the very same system showing the presence of inhomogenities in the local gap structure.[22] Indeed these same SISTM studies indicate the gap distribution is broader in the underdoped region and slowly reduces as the doping increases. The photoemission spectra must reflect this gap distribution. The emitted photoelectrons are not an "average" of the distribution. *Rather, photoelectrons ejected from the different local regions will all be represented by a weighted superposition in the measured spectra.*

In figure 4 we show the results of a simulation of the photoemission spectra that recognizes the gap distribution measured in the spatially resolved SISTM studies.[22] Figure 4(a) shows the gap distributions used in the simulation as a function of doping. As indicated in the supplementary information, comparison to magnetization measurements indicates that the mean field description of the superconducting gap finds a gap pairing strength for these samples at 0K of 2Δ = 5.84T$_c$. Here T$_c$ is clearly defined by the temperature at which overall phase coherence is identified across the entire crystal. The relationship 2Δ = 5.84 kT$_c$, is larger than the standard BCS value of 3.5 kT$_c$,[28] but appropriate we believe for a more strongly coupled gap maximum in a d-wave superconductor. We note that fitting the gap closure in the vicinity of the nodal region also yielded a value of 2Δ = 6.0kT$_c$[29] for the optimally doped material and a similar result was found in Raman studies of the Bi2212 system across the phase diagram.[30]

Figure 4(c) shows the simulation of the measured spectra. In each nano-region the gap is allowed to evolve according to the gap equation given by

$$\Delta(T) = \Delta(0)\tanh[\alpha \left(\frac{T_c}{T} - 1\right)^{\frac{1}{2}}] \tag{1}$$

where α is such that Δ(0) = αkT$_c$.[29] The emission from each region is represented by two Lorenztians separated by the temperature dependent gap. On passing through the T$_c$ for that region the width of each Lorenztian increases by an order of magnitude from 5 meV to approximately 50 meV.[26,27] Note that the simulation shows the overall gap persisting to temperatures above the average T$_c$, measured in the magnetization studies. In figure 4(c) we also show the simulation assuming a single average gap (gray curve) corresponding to the bulk T$_c$. Clearly the gap in the simulation now closes at T$_c$ as would be expected. In figure 4(b) we show the total width of the simulated spectra as a function of temperature for different doping levels. Again the simulations reproduce the measured FWHM of 3(a) in detail. That a gap exists seemingly above T$_c$ in the overdoped regime has often been interpreted as evidence for the continued presence of a pseudogap. However, we believe that the present explanation of local doping or gap variations is more consistent with the results of the SISTM studies of Bi2212 and the Hall coefficient studies of YBCO, referenced earlier, showing no pseudogap for doping levels beyond approximately 0.19.

**b) The Underdoped Regime:**

Having clearly demonstrated that the gap observed above T$_c$ in the overdoped region, beyond the critical doping level, reflects the superconductivity associated with nanoscale inhomogeneities, we now re-examine the pseudogap observed in the underdoped region. We ask the question: is there any relationship between the pseudogap measured in this doping region and the measured superconducting gap, particularly if the former reflects in some way pre-formed pairs? Indeed several early theoretical studies based on the Resonating Valence Bond (RVB) spin liquid model of the cuprates[23] proposed that the spin gap associated with local spin singlet pairing giving rise to a pseudogap at low doping, evolves smoothly into the superconducting gap at higher doping levels.[31,32] Such models combined with other doping dependent phenomena including the role of phase fluctuations[32] and coherence[33] have frequently been invoked to explain the "development" of the superconducting dome. In such models the "pseudogap" line does indeed therefore touch the superconducting dome tangentially. As already discussed, it is not clear that such a concept will hold if there is a critical doping level in the vicinity of 0.19. To examine the properties of the phase diagram further we plot in figure 5 the pseudogap transition temperatures, T*, measured only by ARPES studies on the BSCCO (2212) system.[34,35,36,37] The reason that we make these restrictions is simply that the doping level at which the Fermi surface reconstruction takes place quite often appears to differ from one cuprate family to another.[38] Further, the Bi2212 system has been the subject of the most detailed temperature dependent ARPES studies of the pseudogap and the associated Fermi surface reconstruction. The important observation is that all of these studies provide a measurement of pseudogap closing, an indication of the crossover from the small pockets associated with doped Mott Insulator to the full Fermi surface associated with the "strange" metallic behavior at that particular doping level. We plot $T_c$ as a function of doping level because quite often in different studies, even for the Bi2212 system, the $Tc_{max}$ associated with superconductivity has been reported to range from 90-96K. We may appear to have been selective in the use of data from the study of Vishik et al. That study also included data points recorded from the related Pb doped Bi2212 system. However we note that the latter data points fell outside the pseudogap line defiuned by the authors themselves. Whether this reflects known changes in the BiO layer, increased disorder or a change in the (π-0) line spectra previously reported[39] for the Pb doped system is unclear.

Fitting the measured data points for doping levels between 0.1 and 0.2, the fit shown in the figure extrapolates to a pseudogap temperature T* = 0K at a doping level $x_c =$ 0.193, exactly the doping level at which the SISTM studies indicate a reconstruction from arcs or hole pockets to a full Fermi surface.[17] It is well established that the gap size also shows a linear energy dependence with doping.[40] As such, we can look for a mean-field relationship between the measured temperature, T*, and the measured gap, $\Delta_{PG}$, similar to the relationship derived for the superconducting gap. However at low temperatures it is difficult to distinguish between the pseudogap and the superconducting gap for doping levels less than the critical doping level. We therefore explore the possibility of characterizing the pseudogap by the calculated doping dependent energy scale associated with the formation of the pseudogap.

We have previously shown that the phenomenological YRZ ansatz[15] for the pseudogap regime provides an excellent description of the evolution of the hole pockets with increased doping.[13] The same ansatz is consistent with the crossover from arcs to full Fermi surface observed in the SISTM studies of Bi2212[17] and has also recently been shown to be consistent with the doping dependence observed for the carrier density determined in high magnetic field studies of the Hall coefficient in YBCO.[41] Embedded within the YRZ phenomenology is a self-energy term associated with the doping dependent pseudogap, $\Delta_{PG}$. The latter takes its form from consideration of the exchange interaction term within the framework of a renormalized $t-J$ Hamiltonian associated with the doped Mott insulator,[42,43]

$$H_{eff} = g_t T + g_s J \sum s_i \cdot s_j \qquad (2)$$

Here, $g_s(x) = \frac{4}{(1+x)^2}$, the Gutzwiller factor, reflects the number of pairs of sites that can experience spin exchange at a doping level, $x$. Numerical analysis and subsequent modeling have resulted in the pseudogap energy scale taking the form[15,44]

$$\Delta_{PG} = 0.3t(1 - \frac{x}{x_c}) \qquad (3)$$

where t = 3J with J representing the Heisenberg exchange interaction given by J=4t/U² in the t-J model.

Within this framework setting $CkT^* = 2\Delta_{PG}$, the gradient obtained from the fitting in figure 5 will be given from equation (3) by $-0.6t/Cx_c$. With J = 130 meV, the value obtained from the fit of the YRZ model to arcs/pockets in ref. 13, fits to inelastic scattering of the spin wave spectrum[45], and from Raman spectra[46], we obtain a value of C = 4.27 or $2\Delta_{PG} = 4.27kT^*$. This is almost identical to the value of $2\Delta_{PG} = 4.3kT^*$ found in earlier tunneling spectroscopy studies of a range of cuprates[47] and from Raman studies of the same systems.[30] To turn this statement around, we can note that the mean field description of the pseudogap temperature relationship found in the earlier tunneling spectroscopy studies and the Raman studies *requires the same doping dependent energy scale that drives the observed Fermi surface reconstruction*.[44]

Further, we note that this pseudogap energy scale is derived from short range spin correlations with no reference to long range order in either the spin or charge degrees of freedom.

**Discussion:**

By examining the temperature dependence of the antinodal gap as a function of doping in two distinct regions of the phase diagram we are able to distinguish the pseudogap and the superconducting gap associated with the nanoscale inhomogenities. This analysis leads to a picture in which the pseudogap is associated with singlet formation in the doped Mott insulator. The energy scale associated with the pseudogap is derived from the short range spin correlations but the mean field description indicates a pairing strength smaller than the pairing strength in the superconducting state. It seems very unlikely that the pairing mechanism changes on entering the superconducting phase and as such we assume that the superconducting pairing mechanism also involves spin interactions. Interestingly a recent study of the t – J model[48] found that pairing interaction involving spin excitations increases as the temperature is lowered reflecting a rearrangement in the density of states. The authors of that study note that this would not be the case for interactions involving the lattice.

We further note that the present analysis indicates that the pseudogap line associated with singlet formation penetrates the superconducting dome and intersects the axis at a critical doping level associated with the Fermi surface reconstruction as shown in figure 5. It is important to note that the energy scale which is directly related to 3J takes the same form for both the singlet pairing strength and for the Fermi surface reconstruction. We can ask the question whether or not the singlets should be considered as pre-formed pairs that ultimately condense into the superconducting state or whether their presence actually competes with the superconductivity, which is believed to develop in the nodal region. The observation that with reduced doping the size of the Fermi surface in the nodal region decreases and $T_c$ decreases suggests that it is in fact the former, i.e. the presence of the singlets or pseudogap does compete with the superconducting state.

The data points in figure 5 indicated as references 27 and 34 are recorded in the nodal region in laser based photoemission studies. This is a difficult region of the phase diagram as the gaps associated with true superconductivity are almost identical in magnitude to the gaps associated with the pseudogap. Thus we suggest that laser based photoemission studies are in fact exploring the superconducting state. Several recent photoemission studies have suggested another region in the phase diagram associated with pre-formed pairing and defined a temperature $T_{pair}$, a distinct boundary from the pseudogap line $T^*$[12,27] Indeed we further note that the $T_{pair}$ line or the region defined in one of the latter studies[27] closely parallels the region defined in earlier studies of the Nernst effect.[49] In fact the measured $T^*_2$ as represented in figure 3 tracks this $T_{pair}$ line quite closely. Rather than pre-formed pairing we associate this boundary with the onset of phase coherence or superconducting fluctuations within the larger nano-regions assuming that $T_c$ increases with the length scale associated with those regions.[50] Support for this is given by the observation in one of the studies[27] that on that boundary $2\Delta = 6kT_c$, identical to that reported for the superconducting state here and elsewhere.[29,30] If the latter is true, the pathway to higher transition temperatures in such strongly correlated systems would then be finding methods of engineering these materials in a way that maximizes the size of the nano-regions, associated with disorder or phase separation.

**Acknowledgements:**

The authors acknowledge useful discussions with Maurice Rice, Alexei Tsvelik, John Tranquada, Doug Scalapino, Phuan Ong, Gabi Kotliar, Inna Vishik, Ali Yazdani, and Mike Norman. This work was supported in part by the Center for Emergent Superconductivity (CES), an Energy Frontier Research Center, in part by the Center of Computational Design of Functional Strongly Correlated Materials and Theoretical Spectroscopy, and also in part by the U.S. DOE under Contract No. DE-SC0012704. The work at Argonne is partially supported by the U.S. DOE under Contract No. DE-AC02-06CH11357 and partially by the same CES.

**Figures:**

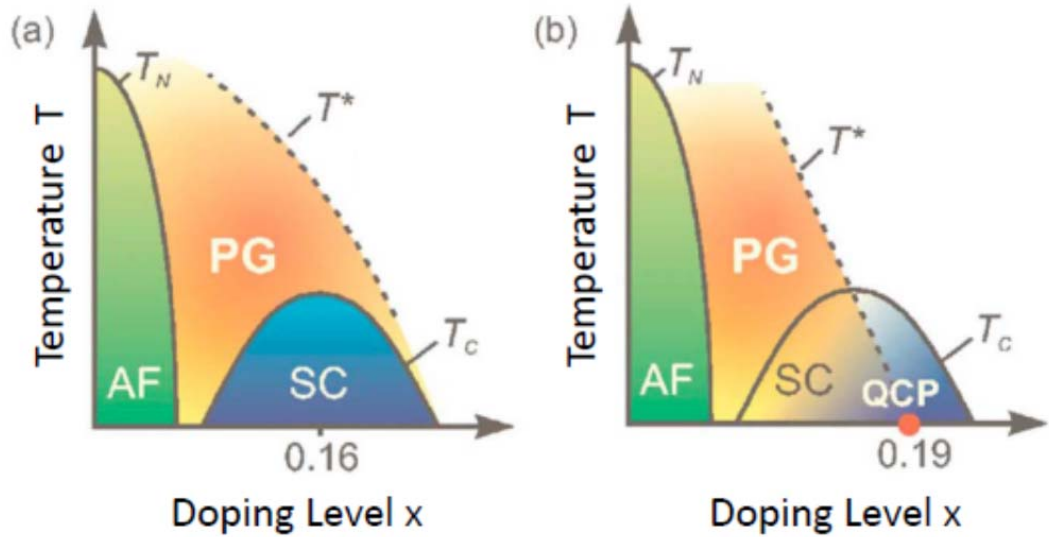

Fig. 1. Two scenarios for the hole-doped HTS phase diagram showing the Antiferromagnetic region (AF), the Pseudogap region (PG) and the Superconducting region (SC). a) T* merges with $T_c$ on the overdoped side. b) T* crosses the superconducting dome and falls to zero at a quantum critical point (QCP). $T_N$ is the Néel temperature for the antiferromagnetic state.

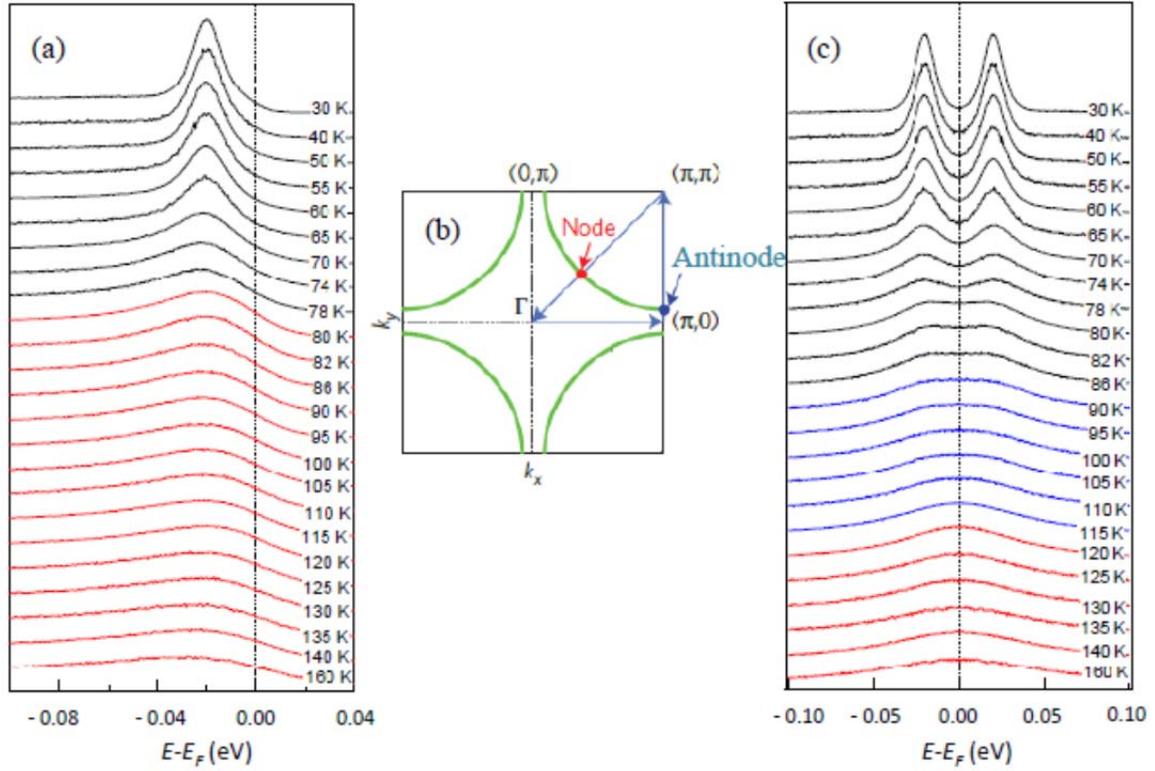

Fig. 2. Photoemission spectra recorded from an overdoped sample with Tc = 80K. a) Temperature dependence of the spectra recorded at the antinodal point indicated in the schematic of the Fermi surface shown in (b). In the latter we indicate both the nodal and antinodal directions. In a) the measured transition temperature $T_c$ is indicated by the transition from black to red curves. (c) The same spectra as in a) symmetrized around the chemical potential. Now the color coding indicates that the gap loses a minimum in the center at approximately 90K but does not finally disappear until approximately 120K.

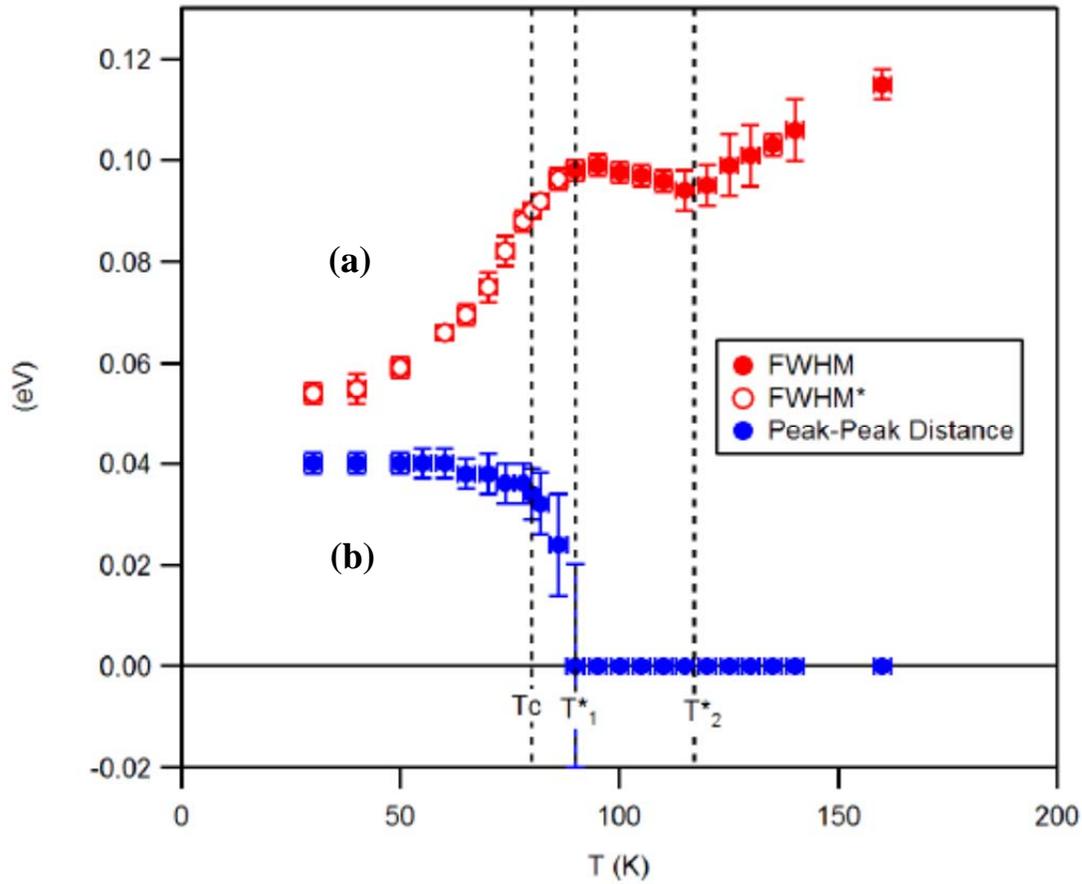

Fig. 3. (a) The FWHM of the two peak structure indicated in figure 2(c). The open circles indicate where two Lorenztians are used to fit the structure, the filled circles indicate a single Lorenztian. (b) The separation of the two Lorenztians used in the fitting procedure. $T^*_1$ indicates the temperature at which there is no longer a dip between the two peaks and $T^*_2$ indicates the temperature at which there is no longer a decrease in the overall width which then simply increases linearly with temperature.

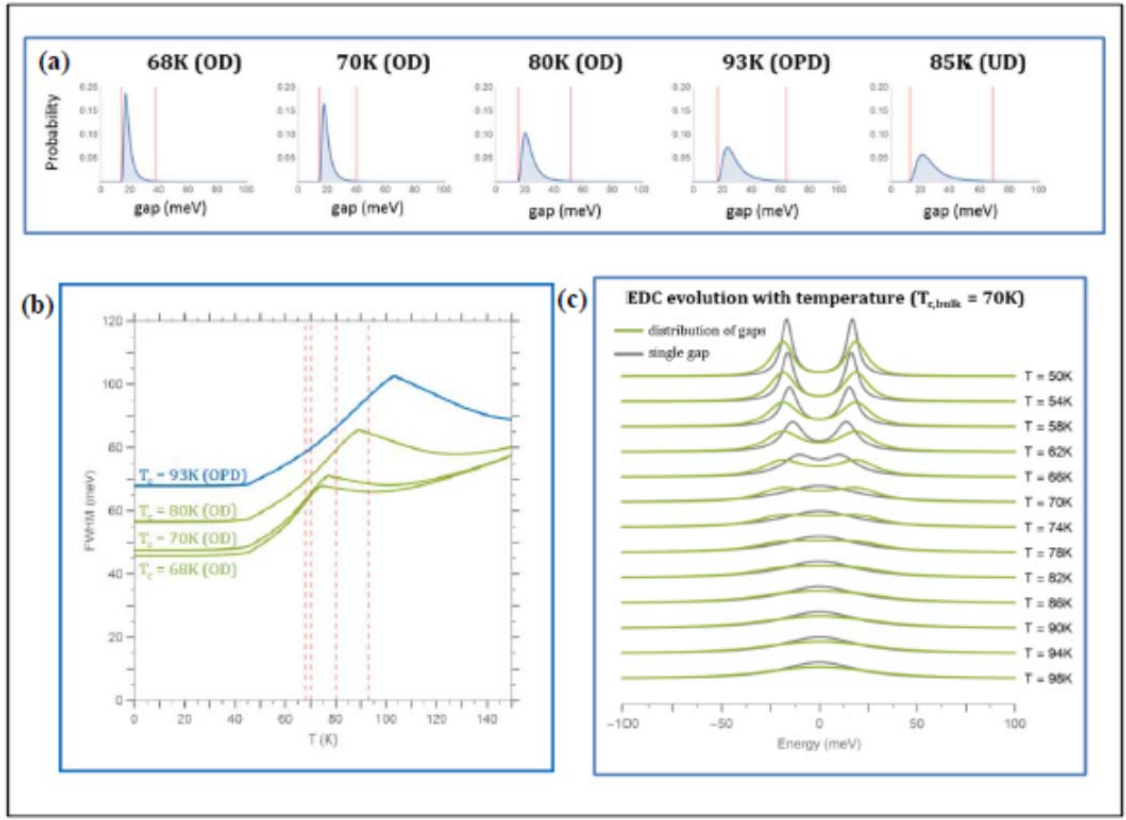

Fig. 4. Simulation of the photoemission spectra in the antinodal region. (a) The gap distributions used in the simulation as a function of doping. The transition temperature Tc and region in the phase diagram is indicated. The distribution peaks at the overall Tc measured for the sample as in magnetization studies. (b) The simulated overall width of the peak structure in the overdoped region as a function of doping to be compared with the experimental measurements as shown in fig. 3(a) for example. (c) Simulation of the measured two peak structure. The gray curve represents a simple two peaked structure associated with the average Tc, the green curve reflects the gap structure associated with the nanoscale inhomogeneities and FWHM indicated in (b).

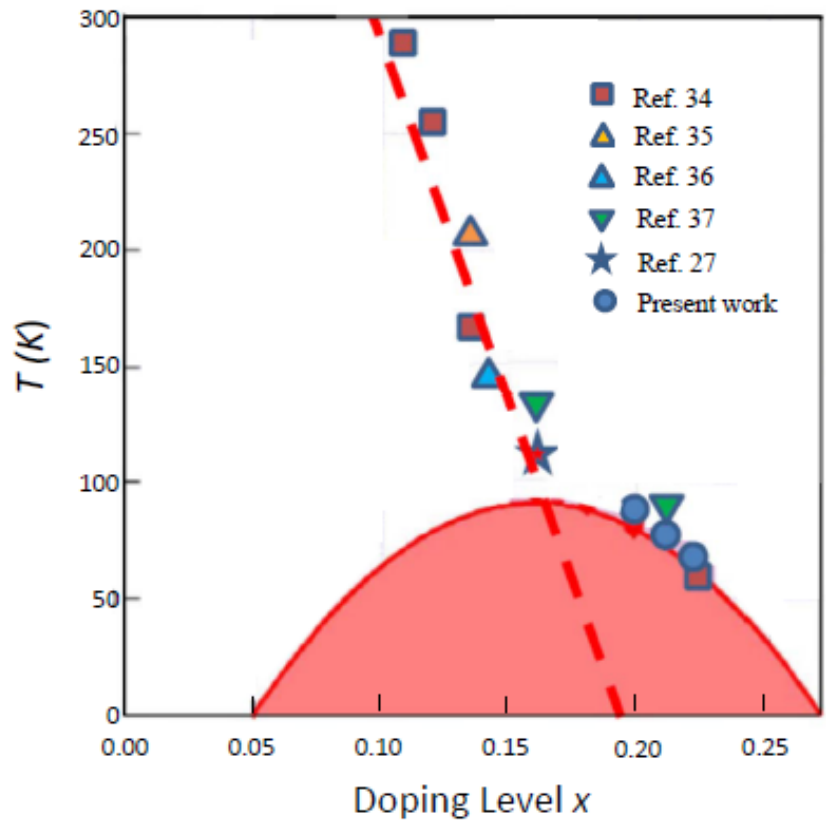

Fig. 5. The pseudogap temperature T* determined in different studies of the Bi2212 system as a function of doping $x$. The different sources are indicated. The red line indicates a fit to the data recorded for doping levels below $x = 0.2$. The references associated with different data points are indicated.

**References:**

1. R. Coleman, B. Giambattista, P. Hansma, A. Johnson, W. McNairy, and C. Slough, Adv. Phys. 37, 559 (1988).
2. R. Comin, A. Frano, M. M. Yee, Y. Yoshida, H. Eisaki, E. Schierle, E. Weschke, R. Sutarto, F. He, A. Soumyanarayanan, Yang He, M. Le Tacon, I. S. Elfimov, Jennifer E. Hoffman, G. A. Sawatzky, B. Keimer, A. Damascelli, *Science* 343, 390, (2014)
3. E. H. da Silva Neto, P. Aynajian, A. Frano, R. Comin, E. Schierle, E. Weschke, A. Gyenis, J. Wen, J. Schneeloch, Z. Xu, S. Ono, G. Gu, M. Le Tacon, A. Yazdani, *Science* 343, 393 (2014)
4. X. M. Chen, V. Thampy, C. Mazzoli, A. M. Barbour, H. Miao, G. D. Gu, Y. Cao, J. M. Tranquada, M. P. M. Dean, and S. B. Wilkins. Phys. Rev. Lett. **117**, 167001, (2016)
5. A. Damascelli. Z. Hussain, and Z-X Shen, Reports on Modern Physics, 75, 473 (2003)
6. A.A. Kordyuk, Low Temperature Physics 41, 319 (2015)
7. H. Ding, T. Yokoya, J.C. Campuzano, T. Takahashi, M. Randeria, M.R. Norman, T. Mochiku, K. Kadowaki, J. Giapintzakis, Nature 382, 51 (1996).
8. A.G. Loeser, Z.X. Shen, D.S. Dessau, D.S. Marshall, C.H. Park, P. Fournier, A. Kapitulnik, Science, 273, 325 (1996).
9. M. Platé, J. D. F. Mottershead, I. S. Elfimov, D. C. Peets, Ruixing Liang, D. A. Bonn, W. N. Hardy, S. Chiuzbaian, M. Falub, M. Shi, L. Patthey, and A. Damascelli, Phys. Rev. Lett. **95**, 077001 (2005)
10. A. Kanigel, M. R. Norman, M. Randeria, U. Chatterjee, S. Souma, A. Kaminski, H. M. Fretwell, S. Rosenkranz, M. Shi, T. Sato, T. Takahashi, Z. Z. Li, H. Raffy, K. Kadowaki, D. Hinks, L. Ozyuzer and J. C. Campuzano, Nature Physics, 2, 447 (2006)
11. H.B. Yang, J.D. Rameau, P.D. Johnson, T. Valla, A. Tsvelik, and G.D. Gu, Nature 456, 77 (2008).
12. A. Kaminski, T. Kondo, T. Takeuchi and G. Gu, Philosophical Magazine, 95, 453, (2015)
13. H.-B. Yang, J. D. Rameau, Z.-H. Pan, G. D. Gu, P. D. Johnson, H. Claus, D. G. Hinks, and T. E. Kidd, Phys. Rev. Lett. **107,** 047003 (2011)
14. X. G. Wen and P. A. Lee, Phys. Rev. Lett. **80**, 2193 (1998).
15. K. Y. Yang, T. M. Rice, and F. C. Zhang, Phys. Rev. B **73**, 174501 (2006).
16. Y. Qi and S. Sachdev, Phys. Rev. B **81**, 115129 (2010).
17. K. Fujita, Chung Koo Kim, Inhee Lee, Jinho Lee, M. H. Hamidian, I. A. Firmo, S. Mukhopadhyay, H. Eisaki, S. Uchida, M. J. Lawler, E.-A. Kim, J. C. Davis, Science, 344, 612 (2014)
18. S. Badoux, W. Tabis, F. Laliberté, G. Grissonnanche, B. Vignolle, D. Vignolles, J. Béard, D. A. Bonn, W. N. Hardy, R. Liang, N. Doiron-Leyraud, Louis Taillefer and Cyril Proust, Nature 531, 210 (2016)
19. S. Benhabib, A. Sacuto, M. Civelli, I. Paul, M. Cazayous, Y. Gallais, M.-A. Méasson, R. D. Zhong, J. Schneeloch, G. D. Gu, D. Colson, and A. Forget, Phys. Rev. Lett. 114, 147001 (2015)
20. L. Mangin-Thro, Y. Sidis, P. Bourges, S. De Almeida-Didry, F. Giovannelli, and I. Laffez-Monot, Phys. Rev. B 89, 094523 (2014)
21. L. Ozyuzer, J. F. Zasadzinski, K.E.Gray, C. Kendziora and N. Miyakawa, Europhysics Letters, 58, 589 (2002)
22. K. K. Gomes, A. N. Pasupathy, A. Pushp, S. Ono, Y. Ando and A. Yazdani, Nature 447, 569, (2007)
23. P. W. Anderson, Science **235**, 1196 (1987)
24. C.M. Varma, P.B. Littlewood, S. Schmitt-Rink, E. Abrahams and A.E. Ruckenstein, Phys. Rev. Lett., 63, 1996 (1989)
25. T. Valla, A.V. Fedorov, P.D. Johnson, B.O. Wells, S.L. Hulbert, Q. Li, G.D. Gu, and N. Koshizuka. Science 285, 2110 (1999).
26. M. R. Norman, M. Randeria, H. Ding, and J. C. Campuzano, Phys. Rev. B **57**, R11093(R) (1998)
27. T. J. Reber, S. Parham, N. C. Plumb, Y. Cao, H. Li, Z. Sun, Q. Wang, H. Iwasawa, J. S. Wen, Z. J. Xu, G. Gu, S. Ono, H. Berger, Y. Yoshida, H. Eisaki, Y. Aiura, G. B. Arnold, D. S. Dessau, arXiv:1509.01556v1 (2015)
28. M. Tinkham, *Introduction to Superconductivity*, Dover Publication, Mineola, NY (2004)
29. T. Dahm, P. J. Hirschfeld, D. J. Scalapino, and L. Zhu, Phys. Rev. B **72**, 214512, 2005.
30. A. Sacuto, Y. Gallais, M. Cazayous, M.-A. Measson, G.D. Gu, and D. Colson, Rep. Prog. Phys., 76, 022502 (2013).
31. G. Kotliar and J. Liu, Phys. Rev. B, **38**, 5142 (1988)
32. V.J. Emery and S.A. Kivelson Nature 374 434 (1995)

[33] H. Ding, J. R. Engelbrecht, Z. Wang, J. C. Campuzano, S.-C. Wang, H.-B. Yang, R. Rogan, T. Takahashi, K. Kadowaki, and D. G. Hinks, Phys. Rev. Lett. **87**, 227001 (2001)

[34] I. M. Vishik, M. Hashimoto, Rui-Hua He, Wei-Sheng Lee, Felix Schmitt, Donghui Lu, R. G. Moore, C. Zhang, W. Meevasana, T. Sasagawa, S. Uchida, Kazuhiro Fujita, S. Ishida, M. Ishikado, Yoshiyuki Yoshida, Hiroshi EisakI, Zahid Hussain, Thomas P. Devereaux, and Zhi-Xun Shen, Proc. Nat. Acad. Sci. 109, 18332 (2012)

[35] Takeshi Kondo, Yoichiro Hamaya, Ari D. PalczewskI, Tsunehiro TakeuchI, J. S. Wen, Z. J. Xu, Genda Gu, Jörg Schmalian, and Adam Kaminski, Nat. Phys. 7, 21 (2011)

[36] A. Kaminski, S. Rosenkranz, H. M. Fretwell, J. C. Campuzano, Z. Li, H. Raffy, W. G. Cullen, H. You, C. G. Olson, C. M. Varma and H. Höchst, Nature 416, 610 (2002)

[37] Takeshi Kondo, W. Malaeb, Y. Ishida, T. Sasagawa, H. Sakamoto, Tsunehiro TakeuchI, T. Tohyama, and S. Shin, Nat. Comm., 6, 7699 (2015)

[38] For instance in a study of the single layer BSCCO (2201) system, Y. He et al., Science, 344, 608 (2014) suggest that the critical point is at a doping level of 0.15.

[39] S.V. Borisenko, T.K. Kim, A.A. Kordyuk, M. Knupfer, J. Fink, J.E. Gayone, P. Hofmann, H. Berger, B. Liang, A. Maljuk, C.T. Lin, Physica C 417, 1-6 (2004)

[40] S Hufner, M A Hossain, A Damascelli and G A Sawatzky, Rep. Prog. Phys. 71, 062501 (2008)

[41] J.G. Storey, EPL, 113, 27003 (2016)

[42] F C Zhang, C Gros, T M Rice and H Shiba, Supercond. Sci. Tech., 1, 36 (1988)

[43] P.W. Anderson, P.A. Lee, M. Randeria, T.M. Rice, N. Trivedi and F.C. Zhang, Journal of Physics: Condens. Matt. 16, 755 (2004)

[44] K.-Y. Yang, H.-B. Yang, P.D. Johnson, T.M. Rice, and F.-C. Zhang, Europhys. Lett. **86**, 37 002 (2009).

[45] R. Coldea, S. M. Hayden, G. Aeppli, T. G. Perring, C. D. Frost, T. E. Mason, S.-W. Cheong, and Z. Fisk, Phys. Rev. Lett., 86, 5377 (2001)

[46] S. Sugai, H. Suzuki, Y. Takayanagi, T. Hosokawa, and N. Hayamizu, Phys. Rev. B 68, 184504 (2003)

[47] Ø. Fischer, M. Kugler, I. Maggio-Aprile, C. Berthod, and C. Renner, Rev. Mod. Phys. 79, 353, (2007)

[48] T.A. Maier, P. Staar, V. Mishra, U. Chatterjee, J.C. Campuzano and D.J. Scalapino, Nat. Comm. 7, 11875, (2016)

[49] Y. Wang, L. Li, and N. P. Ong, Phys. Rev. B 73, 024510 (2006).

[50] D.S. Fisher, Physica A, 263, 222 (1999)

# Supplementary Information: The cuprate phase diagram and the influence of nanoscale inhomogeneities.

## 1. Crystal synthesis and characterization:

All of the cuprate overdoped crystals used in the present study were grown using the floating-zone method, and their $T_c$ values were adjusted by both oxygen annealing and cat-ion doping. The midpoint of the superconducting transition in the magnetic susceptibility curve was defined as the $T_c$ value. Typical magnetization curves are shown in Fig. S1.

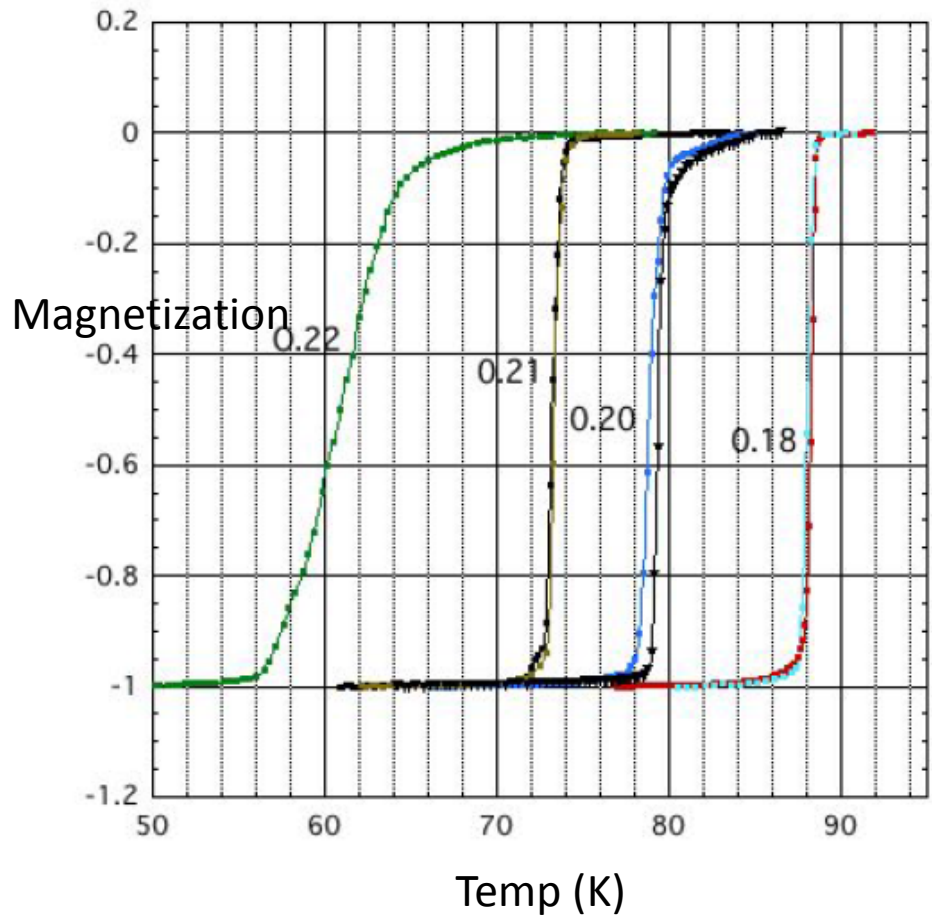

Fig. S1. Typical magnetization curves obtained from the crystals used in the present study. The indicated doping levels are derived via the standard equation from the measured $T_c$'s.

Figure S2 shows the gap measured in the anti-nodal direction at low temperatures as a function of $T_c$ measured in the magnetization studies.

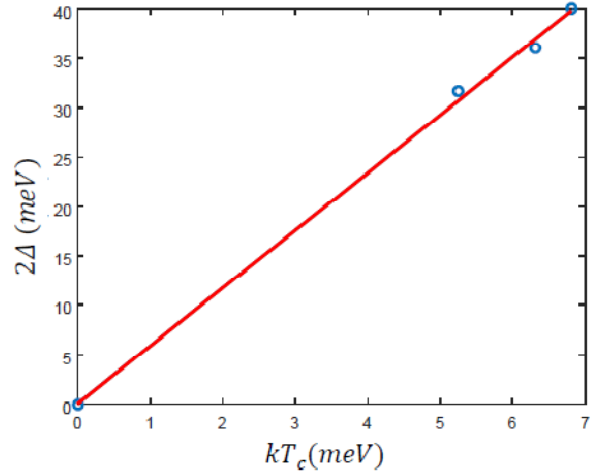

Fig. S2. Plot of the gap measured in the antinodal direction at low temperatures as a function of $T_c$ determined in magnetization measurements.

## 2. ARPES spectrum analysis:

Figure S3 shows the raw data as measured at the anti-nodal point indicated in figure 2(b) in the main text and the same data after both symmetrizing and normalizing by the Fermi-Dirac Distribution. It can be seen that the latter two analyses provide nearly identical information. Figure S4 shows Lorentzian fits to the normalized raw data.

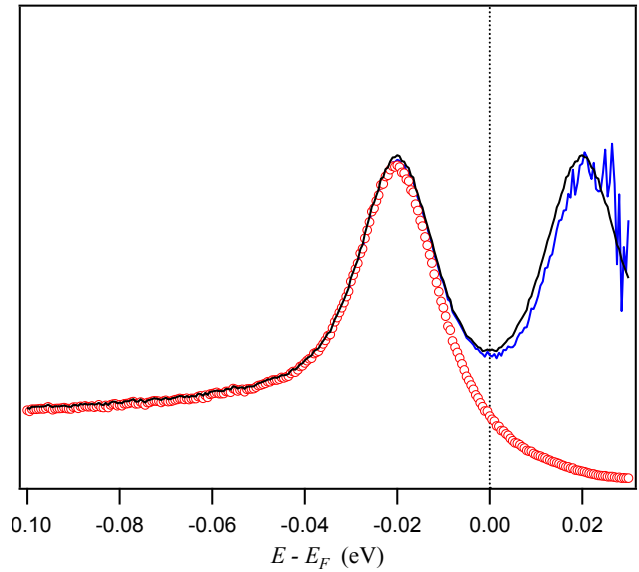

Fig. S3. A comparison of the fitted spectrum (black line) showing the complete gap obtained by symmetrizing the raw data (red circles) and normalizing to the Fermi-Dirac distribution, with the spectrum (blue line) obtained solely by normalizing the raw data to the Fermi-Dirac distribution.

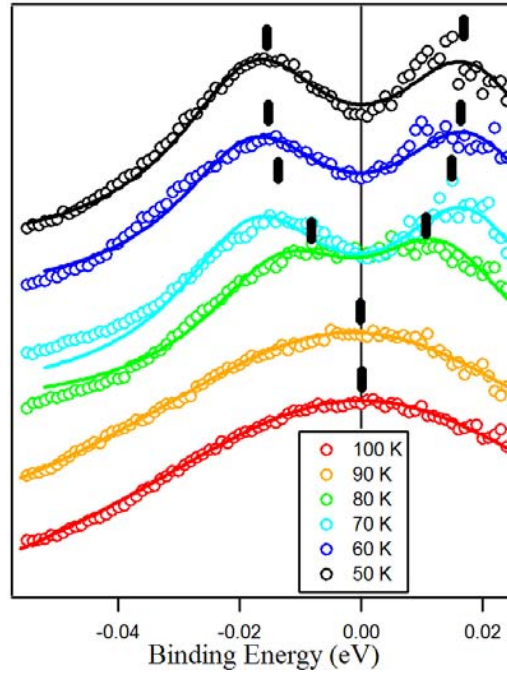

Fig. S4. Representative Lorentzian fits to the symmetrized raw data normalized to the Fermi-Dirac distribution function.

Data obtained from fitting the temperature dependent gap distributions in the anti-nodal direction for different doping levels is shown in figure S5. The figure also indicates the magnitude of the gap as a function of temperature.

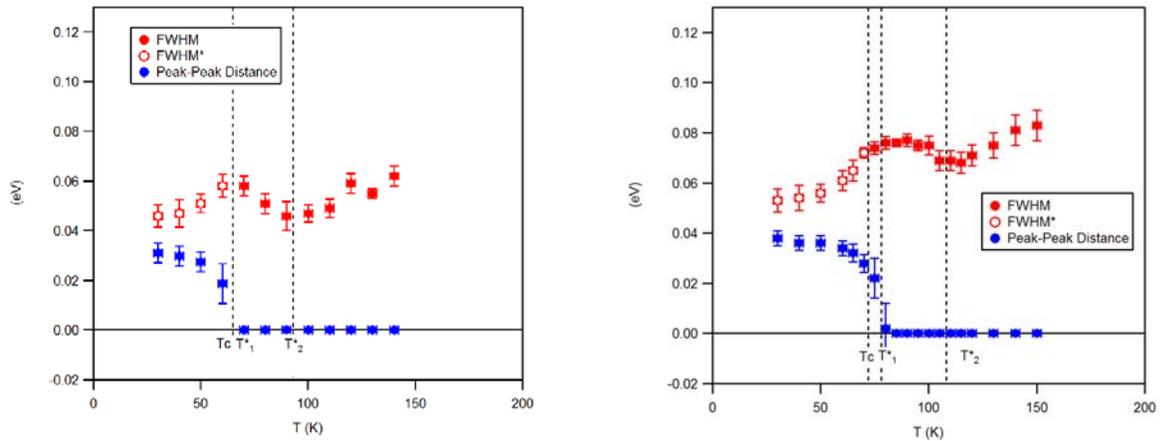

Fig. S5. The FWHM of the two peak fitting to the gap structure for different doping levels. The open circles indicate where two Lorentzians are used to fit the structure, the filled circles indicate a single Lorentzian. The separation of the two Lorentzians used in the fitting procedure is also indicated (in blue). $T^*_1$ indicates the temperature at which there is no longer a dip between the two peaks and $T^*_2$ indicates the temperature at which there is no longer a decrease in the overall width which then simply increases linearly with temperature. (a) Presents data from a Tc = 65K sample and (b) presents data from a Tc = 70K sample.

## 3. Gap simulation:

The photoemission gap spectrum at the anti-nodal region was simulated using a sum of Lorentzian pairs. The peak-to-peak separation of each Lorentzian pair is equal to their respective gap size ($\Delta$), and evolves with temperature using the BCS gap equation:

$$\Delta(T) = \Delta(0)\, tanh\,[\alpha\,(T_c/T - 1)^{1/2}] \text{ for } 0 \leq T \leq T_c, \text{ and}$$
$$\Delta(T) = 0 \text{ for } T > T_c$$

where $\Delta(0) = \alpha k T_c$, and $\alpha = 2.93$. The linewidth ($\omega$) of the Lorentzian pairs was also allowed to evolve with temperature using a similar equation:

$$\omega(T) = (\Delta(T)/\Delta(0))\,\omega(0) + (1 - (\Delta(T)/\Delta(0)))\,\omega(T_c) \text{ for } 0 \leq T \leq T_c, \text{ and}$$
$$\omega(T) = \omega(T_c) \text{ for } T > T_c$$

where $\omega(0) = 5$ meV and $\omega(T_c) = 46$ meV. To account for the Marginal Fermi liquid behavior in the normal state a linearly increasing temperature dependent broadening was added to the linewidth of the Lorentzian pairs. The peak-to-peak separation and relative amplitudes of each Lorentzian pair was based on the STM derived gap distribution for the particular doping of interest [S1]. For the purpose of the simulation of the STM derived gap distributions [S1] it was found that a generalized Planck distribution, also known as a Davis distribution [S2] was a perfectly reasonable starting point.

References:

[S1] Gomes et al, Nature **447**, 569 (2007)
[S2] Christian Kleiber, Samuel Kotz, *Statistical Size Distributions in Economics and Actuarial Sciences*, Wiley (2003)